\documentclass[conference]{IEEEtran}
\IEEEoverridecommandlockouts

\usepackage{cite}
\usepackage{amsmath,amssymb,amsfonts}
\usepackage{graphicx}
\usepackage{textcomp}
\usepackage{xcolor}
\usepackage{url}

\usepackage{booktabs}   
\usepackage{multirow}
\usepackage{array}
\usepackage{tabularx}
\usepackage{float}
\usepackage{colortbl}
\usepackage{caption}
\usepackage{adjustbox}

\usepackage{algorithm}
\usepackage{algorithmic}

\usepackage{fancyhdr}

\usepackage{hyperref}
\hypersetup{
    colorlinks=true,
    citecolor=blue,
    linkcolor=black,
    urlcolor=blue
}
\def\BibTeX{{\rm B\kern-.05em{\sc i\kern-.025em b}\kern-.08em
T\kern-.1667em\lower.7ex\hbox{E}\kern-.125emX}}

\begin{document}

\title{CANGuard: A Spatio‑Temporal CNN‑GRU‑Attention Hybrid Architecture for Intrusion Detection in In‑Vehicle CAN Networks\\
}

\author{
    \IEEEauthorblockN{
        Rakib Hossain Sajib\textsuperscript{1}, 
        Md. Rokon Mia\textsuperscript{2}, 
        Prodip Kumar Sarker\textsuperscript{3}, 
        Abdullah Al Noman\textsuperscript{4},\\
        Md Arifur Rahman\textsuperscript{5}
    } 
    \IEEEauthorblockA{
        \textsuperscript{1,2,3}Department of Computer Science and Engineering, Begum Rokeya University, Rangpur, Rangpur, Bangladesh  \\
        \textsuperscript{4}Wilmington University, New Castle, Delaware, United States \\
        \textsuperscript{5}Trine University, Angola, Indiana, United States\\
        Email: rakibnsajib@gmail.com,
        miarokon2001@gmail.com, 
        prodip@brur.ac.bd,\\
        anoman001@my.wilmu.edu,
        mrahman22@my.trine.edu
    }
}

\maketitle
\thispagestyle{fancy} 
\fancyhf{} 

\renewcommand{\headrulewidth}{0pt}
\renewcommand{\footrulewidth}{0pt}

\begin{abstract}
The Internet of Vehicles (IoV) has become an essential component of smart transportation systems, enabling seamless interaction among vehicles and infrastructure. In recent years, it has played a progressively significant role in enhancing mobility, safety, and transportation efficiency. However, this connectivity introduces severe security vulnerabilities, particularly Denial-of-Service (DoS) and spoofing attacks targeting the Controller Area Network (CAN) bus, which could severely inhibit communication between the critical components of a vehicle, leading to system malfunctions, loss of control, or even endangering passengers' safety.  To address this problem, this paper presents CANGuard, a novel spatio-temporal deep learning architecture that combines Convolutional Neural Networks (CNN), Gated Recurrent Units (GRU), and an attention mechanism to effectively identify such attacks. The model is trained and evaluated on the CICIoV2024 dataset, achieving competitive performance across accuracy, precision, recall, and F1-score and outperforming existing state-of-the-art methods. A comprehensive ablation study confirms the individual and combined contributions of the CNN, GRU, and attention components. Additionally, a SHAP analysis is conducted to interpret the decision-making process of the model and determine which features have the most significant impact on intrusion detection. The proposed approach demonstrates strong potential for practical and scalable security enhancements in modern IoV environments, thereby ensuring safer and more secure CAN bus communications.

\end{abstract}


\section{Introduction}
The Internet of Vehicles (IoV) is an advancement of traditional Vehicle Ad-hoc Networks (VANETs) into a more integrated network framework. It integrates vehicles, humans, things, and environments into a global network, enabling various services and applications \cite{6969789}. It has emerged as a vital component in the progression of smart vehicle transportation systems and facilitates smooth interaction between vehicles and infrastructure \cite{Contreras-Castillo2018Internet}. As IoV systems become increasingly interconnected, they face heightened cybersecurity threats \cite{morol2022data}, especially Denial-of-Service (DoS) and spoofing attacks aimed at the Controller Area Network (CAN) bus.

The CAN bus is a standardized communication protocol designed for in-vehicle data exchange among Electronic Control Units (ECUs) \cite{Tuohy2015Intra-Vehicle}. It facilitates high-speed real-time communication for safety-critical functions like braking systems, engine management, and airbag deployment, while also supporting slower communication for air-conditioning, instrument clusters, and wiper control \cite{Ye2010Research}. However, it is deficient in built-in security functionalities, including authentication and encryption, making it highly vulnerable to cyberattacks. DoS attacks can flood the CAN bus with excessive messages, resulting in communication breakdowns and vehicle malfunctions. In contrast, spoofing attacks inject malicious data to impersonate Electronic Control Units, thereby altering vehicle functionality and posing significant safety hazards, such as loss of control, unintended acceleration, or brake failure \cite{10.1145/3542954.3543028}.

The increasing prevalence of cyber attacks has emphasized the necessity for robust safety protocols inside IoV systems in recent years. To tackle these concerns, this paper presents a hybrid CNN architecture specifically designed to detect DoS and spoofing threats on the CAN bus within IoV environments. The key contributions of this study are outlined below:
\begin{itemize}
    \item CANGuard, a spatio-temporal CNN–GRU–Attention intrusion detection model specifically optimized for CAN bus traffic in Internet of Vehicles (IoV) environments.
    \item Sequence-aware representation of CAN messages, capturing temporal payload dependencies in in-vehicle CAN networks.
    \item Component-wise ablation study, quantifying the individual and combined impact of CNN, GRU, and attention modules on multi-class CAN intrusion detection.
    \item CAN-payload-level interpretability using SHAP, highlighting the contribution of individual data bytes (DATA\_0–DATA\_7) to the detection of DoS and spoofing attacks.
    \item A strong empirical baseline on the CICIoV2024 dataset for multi-class intrusion detection in in-vehicle CAN networks.
\end{itemize}

\section{Literature Review}
\label{sec:related_work}
In response to the growing cybersecurity challenges in IoV, a variety of studies have explored the application of machine learning for enhancing the security of in-vehicle networks through Intrusion Detection Systems (IDS). These studies focus on developing effective machine learning models proficient in accurately identifying attacks on the CAN bus.
Neto et al. \cite{Neto2024CICIoV2024:} employed several machine learning techniques, including Logistic Regression, Random Forest, AdaBoost, and Deep Neural Networks (DNN). About an accuracy of 96\%, the DNN model proved better than all others.  Narkedimilli et al. \cite{Narkedimilli2025Enhancing} proposed a lightweight curriculum-learning architecture enhanced with Explainable AI (XAI), achieving 98\% accuracy. Anargya et al. \cite{Anargya2025Random} applied Random Under-Sampling (RUS) on the CICIoV2024 dataset and showed that Random Forest outperformed Decision Tree and KNN with 98.5\% accuracy and F1-score.

In the broader context of anomaly detection, Lin et al.~\cite{Lin2019} presented a temporal sequence learning model utilizing Long Short-Term Memory (LSTM) networks combined with an attention mechanism. Using SMOTE for data augmentation, their framework attained a detection accuracy of 96.2\%. Comparative studies by Hai and Nam~\cite{Hai2021}, Meliboev et al.~\cite{Meliboev2022}, and Ravinder Reddy et al.~\cite{RavinderReddy2021} evaluated the performance of RNN, LSTM and Gated Recurrent Unit (GRU) architectures on standard datasets such as CSE-CIC-IDS2018 and CICIDS2017. Hai and Nam et al.~\cite{Hai2021} reported nearly equivalent performance across all models, with accuracies approaching 93.9\%. Soltani et al.~\cite{Soltani2023A} developed a multi-agent adaptive deep learning framework for intrusion detection, addressing concept drift in network traffic and the complexities of distributed systems. The researchers utilized two deep learning methodologies: CNN and LSTM networks. Their methodology attained an accuracy of 95\%, signifying commendable performance. Henry et al.~\cite{Henry2023Composition} aimed to improve network intrusion detection by the utilization of a hybrid deep learning framework integrated with feature optimization techniques. Their methodology combined CNN with GRU to enhance network parameter analysis, thus augmenting the detection of anomalies in network traffic statistics. The proposed model attained a remarkable detection accuracy of 98.73\%. The research reveals that deep learning-based methods are exceptionally effective for classifying network behaviors.

The security of the Internet of Vehicles (IoV) is an essential task that directly influences passenger safety, underscoring the necessity for enhanced Intrusion Detection System (IDS) models. Despite significant advancements, existing approaches face challenges such as limited temporal modeling capabilities, insufficient attention to critical features, and higher false positive rates, restricting their applicability in real-world contexts.

In this work, we propose a novel CNN-GRU-Attention hybrid deep learning framework for intrusion detection, leveraging the complementary strengths of Convolutional Neural Networks for spatial feature extraction, Gated Recurrent Units for temporal pattern recognition, and attention mechanisms for enhanced feature focus. Through comprehensive ablation experiments, we demonstrate the contribution of each architectural component and optimize the model for superior performance. The proposed framework significantly surpasses existing state-of-the-art models, improving the security of CAN bus networks in connected vehicles while providing explainable insights through attention visualization.

\section{Methodology}
\label{sec:methodology}
\begin{figure*}
    \centering
    \includegraphics[width=0.95\linewidth]{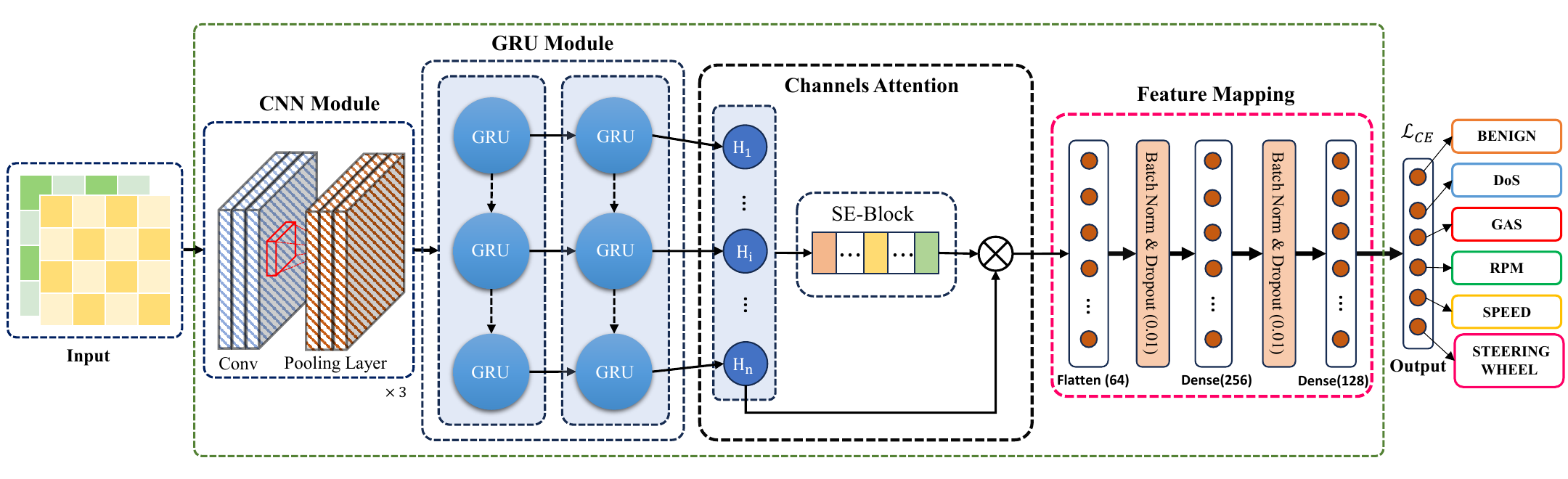}
    \caption{CNN-GRU-Attention architecture employs CNN layers to extract spatial features, GRU layers to model temporal dependencies, an attention mechanism to emphasize important features, and dense layers to perform the final classification.}
    \label{fig:architecture-figure}
\end{figure*}

This research employs a systematic approach to design and evaluate a novel CNN-GRU-Attention architecture for Intrusion Detection Systems (IDS) in IoV environments. The overall methodology encompasses dataset preprocessing, sequence-based feature engineering, deep learning model architecture design with attention mechanisms, comprehensive ablation studies, and performance evaluation against state-of-the-art methods. The complete process flow for developing the CNN-GRU-Attention IDS is illustrated in Fig. \ref{fig:architecture-figure}, and the detailed algorithmic steps are outlined in Algorithm~\ref{alg:CNN_GRU_Attn}.

\subsection{Data Collection}
The CICIoV2024 dataset was used for training and evaluating the proposed IDS in CAN bus. The dataset contains 1,408,219 samples and 12 features, containing real-world IoV traffic data, which includes DoS and spoofing threats targeting the CAN bus, as well as normal traffic data \cite{Neto2024CICIoV2024:}. The dataset consists of six unique categories:
\begin{itemize}
    \item \textbf{BENIGN}: Normal traffic.
    \item \textbf{DoS}: Denial-of-Service attacks.
    \item \textbf {GAS, RPM, SPEED, STEERING\_WHEEL}: Spoofing attacks targeting specific vehicular components.
\end{itemize}
\subsection{Data Preprocessing}
The data set was subjected to several preprocessing steps to assure its appropriateness for machine learning algorithms. The steps involved were:
\subsubsection{Handling Duplicate Values} The dataset was checked for duplicate entries. All the duplicate rows were identified and removed to avoid redundancy.
\subsubsection{Feature Selection} There were certain columns, including `ID', `category', and `label', were excluded from the feature set as they were not directly contributing to detection.
\subsubsection{Temporal Sequence Construction}
To model temporal dependencies in CAN bus data, sliding window sequences were generated. Each sequence consists of $T$ consecutive time steps. The sequence construction process can be formally defined as:

\begin{equation}
X_{\text{seq}} = \left\{ \left( x_{i:i+T},\ y_{i+T-1} \right) \right\}_{i=1}^{N-T+1}
\end{equation}

where $x_{i:i+T}$ is a sequence of $T$ consecutive feature vectors, $y_{i+T-1}$ is the label of the last time step, $T$ is the sequence length, and $N$  indicates the total number of instances.

This transformation converts the original dataset of $N$ instances into $N - T + 1$ overlapping temporal sequences, each shaped as $(T \times F)$, wherein $F$ denotes the number of features.

\subsubsection{Handling Class Imbalance}
The dataset exhibited severe class imbalance. To mitigate this, BorderlineSMOTE was applied to synthetically generate samples for minority classes. This was done by flattening the temporal data before oversampling and then reshaping it back:
\begin{equation}
X^{\text{flat}} \in \mathbb{R}^{N \times (T \cdot F)} \rightarrow \text{BorderlineSMOTE} \rightarrow X' \in \mathbb{R}^{N' \times T \times F}
\end{equation}
Extremely underrepresented classes were retained with class weights $\omega_c$ computed as:
\begin{equation}
\omega_c = \frac{N_{\text{total}}}{N_c}
\end{equation}
where $N_c$ refers to how many samples belong to class $c$.

\subsubsection{Feature Scaling}
Each feature was normalized using Z-score standardization:
\begin{equation}
x' = \frac{x - \mu}{\sigma}
\end{equation}
where $\mu$ denotes the mean value and $\sigma$ stands for the standard deviation. This ensured that all features had zero mean and unit variance, which is essential for the stable convergence of gradient-based learning algorithms, particularly CNNs and RNNs.

\begin{algorithm}
\caption{CNN-GRU-Attention Architecture}
\begin{algorithmic}[1]
\STATE \textbf{Input:} Sequence data \( X_{\text{seq}} \in \mathbb{R}^{N \times T \times F} \)
\STATE \textbf{Output:} Class prediction \( \hat{y} \in \{0, 1, 2, 3, 4, 5\} \)

\STATE \textbf{CNN Feature Extraction:}
\begin{align}
h_{\text{CNN}} =\ & \text{Dropout} \big( \text{MaxPool} \big( \text{Conv1D}_3 ( \nonumber \\
& \text{Conv1D}_2 (\text{Conv1D}_1 (X_{\text{seq}}) ) \big) \big) \nonumber
\end{align}

\STATE \textbf{Bidirectional GRU Processing:}
\[
h_{\text{GRU}} = \text{StackedBiGRU}(h_{\text{CNN}})
\]

\STATE \textbf{Attention Mechanism:}
\[
e_t = \tanh(W_a h_{\text{GRU},t} + b_a), \quad 
\alpha_t = \frac{\exp(e_t)}{\sum_{k=1}^T \exp(e_k)}
\]
\[
c = \sum_{t=1}^T \alpha_t h_{\text{GRU},t}
\]

\STATE \textbf{Classification:}
\[
h_{\text{dense}} = \text{Dropout}(\text{ReLU}(W_1 c + b_1))
\]
\[
\hat{y} = \text{softmax}(W_2 h_{\text{dense}} + b_2)
\]
\end{algorithmic}
\label{alg:CNN_GRU_Attn}
\end{algorithm}

\subsection{CNN-GRU-Attention Architecture}
The proposed CNN-GRU-Attention model integrates three complementary deep learning components to achieve superior intrusion detection performance. The architecture is designed to learn both spatial and temporal features from CAN bus data while providing interpretable insights through attention mechanisms.

\subsubsection{Convolutional Neural Network (CNN) Component}
The CNN component serves as the spatial feature extractor and consists of three sequential Conv1D layers with 64, 128, and 256 filters (kernel size 3, ReLU activation). Each layer is followed by batch normalization, max pooling with a pool size of 2, and a dropout rate of 0.3 for regularization.

\subsubsection{Gated Recurrent Unit (GRU) Component}
The GRU module learns temporal relationships in the CAN bus sequences using two stacked bidirectional GRU layers with 128 and 64 units, respectively, each configured with a dropout rate of 0.3 and recurrent dropout of 0.3. The bidirectional design enables the model to take advantage of both preceding and subsequent contexts within each sequence.

\subsubsection{Attention Mechanism}
The attention mechanism enhances model interpretability and performance by learning to concentrate on the most important temporal features \cite{bhadra2025dual}:
\begin{equation}
\alpha_t = \frac{\exp(e_t)}{\sum_{k=1}^T \exp(e_k)}
\end{equation}
where $e_t = \tanh(W_a h_t + b_a)$ and $h_t$ represents the GRU output at a specific time step $t$.

The resulting context vector after applying attention is computed as:
\begin{equation}
c = \sum_{t=1}^T \alpha_t h_t
\end{equation}

\begin{figure}[H]
    \centering
    \includegraphics[width=1\linewidth]{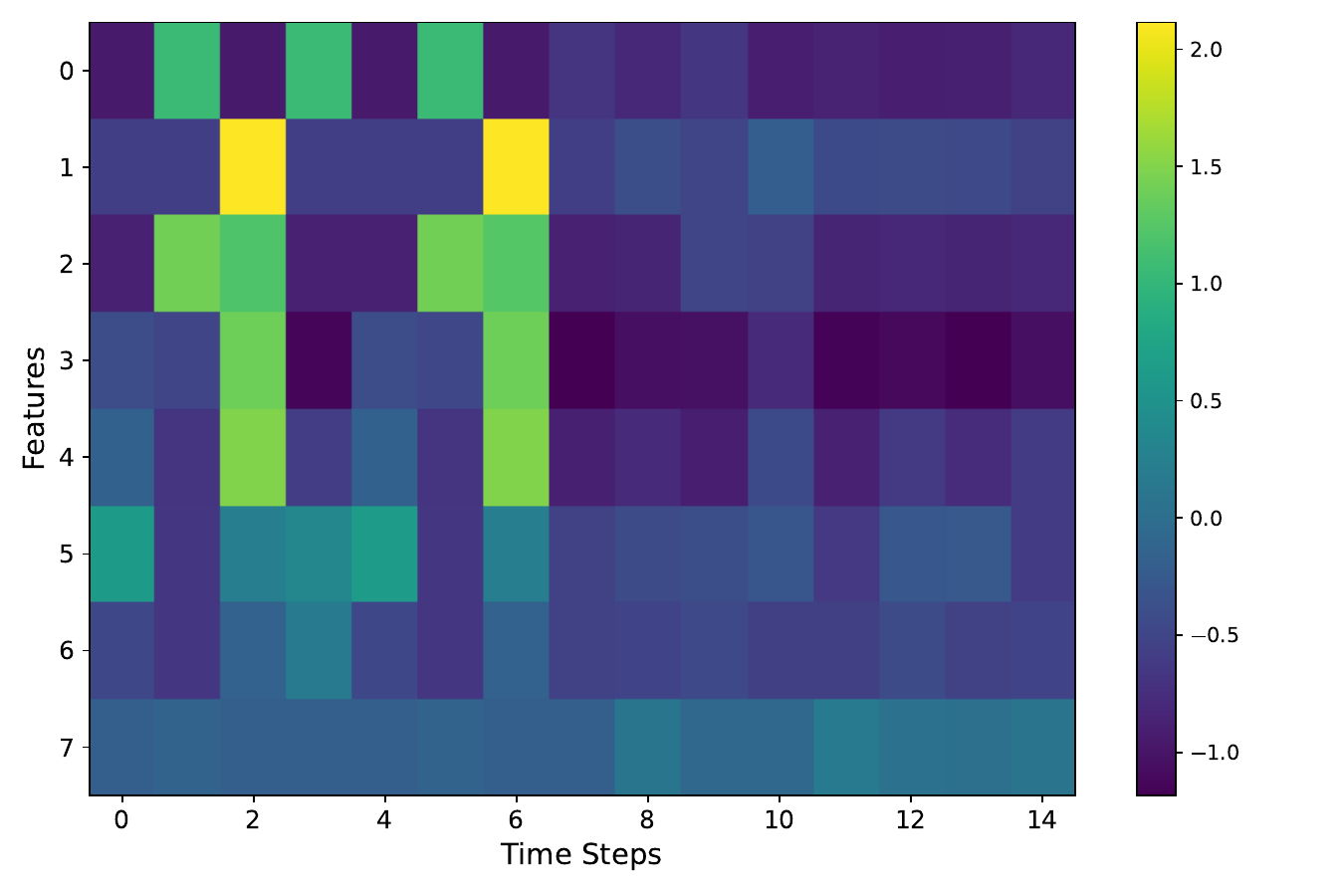}
    \caption{Visual heatmap representation of the attention mechanism's focus across time steps and features, with color intensity indicating the level of importance.}
    \label{fig:attention-heatmap}
\end{figure}

\subsubsection{Classification Layers}
The final classification is performed by two fully connected layers with 256 and 128 neurons (ReLU activation, with L2 regularization and dropout of 0.3 in the second layer), followed by a softmax output layer with 6 neurons corresponding to the target classes.

\subsection{Model Training and Optimization}
The CNN-GRU-Attention model was trained using advanced optimization techniques, such as early stopping, learning rate adjustment, and checkpoint saving. The training process utilized GPU acceleration for efficient computation while maintaining numerical stability through batch normalization and gradient clipping. The detailed hyperparameters and training setup are summarized in Table~\ref{table:training-config}.

\begin{table}[h]
\centering
\caption{Training Configuration}
\label{table:training-config}
\renewcommand{\arraystretch}{1.2}
\begin{tabular}{l|l}
\hline
\textbf{Parameter}       & \textbf{Value} \\
\hline
Optimizer                & Adam with learning rate 0.001 \\
Loss Function            & Categorical crossentropy \\
Batch Size               & 64 \\
Epochs                   & 50 with early stopping (patience = 10) \\
Regularization           & Dropout (0.3), L2 regularization ($\lambda$ = 0.001) \\
\hline
\end{tabular}
\end{table}

The model training process incorporated class weight balancing to handle dataset imbalance and SMOTE augmentation for minority class enhancement. The bidirectional GRU layers enable the model to capture both forward and backward temporal dependencies, while the attention mechanism provides interpretable feature importance scores.

\subsection{Evaluation Metrics}
To evaluate our models' performance, we selected a collection of commonly accepted evaluation metrics. The equations of evaluation metrics are as follows:

\begin{align}
\text{Accuracy} &= \frac{\text{True}^{+} + \text{True}^{-}}%
{\text{True}^{+} + \text{True}^{-} + \text{False}^{+} + \text{False}^{-}}\\
\text{Precision} &= \frac{\text{True}^{+}}%
{\text{True}^{+} + \text{False}^{+}}\\
\text{Recall}    &= \frac{\text{True}^{+}}%
{\text{True}^{+} + \text{False}^{-}}\\
F_{1}\text{ Score} &= 2 \times \frac{\text{Precision} \times \text{Recall}}%
{\text{Precision} + \text{Recall}}
\end{align}

Where $\text{True}^{+}=\text{True Positive},\ \text{True}^{-}=\text{True Negative},\ \text{False}^{+}=\text{False Positive},\ \text{False}^{-}=\text{False Negative}.$

\section{Results and Discussion}

\begin{table*}
\centering
\caption{Comparative Analysis with State-of-the-Art}
\label{table:comparison-sota}
\renewcommand{\arraystretch}{1.2}
\begin{tabularx}{\textwidth}{
  >{\raggedright\arraybackslash}p{2.6cm}  
  >{\centering\arraybackslash}p{0.8cm}    
  >{\raggedright\arraybackslash}p{3.6cm}  
  >{\raggedright\arraybackslash}p{2.5cm}  
  >{\centering\arraybackslash}p{1.35cm}      
  >{\centering\arraybackslash}p{1.35cm}      
  >{\centering\arraybackslash}p{1cm}    
  >{\centering\arraybackslash}p{1.6cm}    
}
\toprule
\textbf{Reference} & \textbf{Year} & \textbf{Method} & \textbf{Dataset} & \textbf{Accuracy(\%)} & \textbf{Precision(\%)} & \textbf{Recall(\%)} & \textbf{F1-Score(\%)} \\
\midrule
\multirow{3}{=}{Neto et al. \cite{Neto2024CICIoV2024:}} & \multirow{3}{*}{2024} 
& Logistic Regression          & CICIoV2024         & 89 & 48  & 50 & 49 \\
 &  & Deep Neural Network          & CICIoV2024         & 96 & 76  & 83 & 78 \\
 &  & AdaBoost                     & CICIoV2024         & 92 & 48  & 66 & 51 \\
 Narkedimilli et al. \cite{Narkedimilli2025Enhancing} & 2025 & Adaptive Curriculum Learning & Custom CAN Dataset & 98 & \textbf{100} & 97 & 97 \\
Anargya et al. \cite{Anargya2025Random}              & 2025 & Random Forest                & Custom IDS Dataset & 98.5 & 98.6 & 98.5 & 98.5 \\
LaxmiLydia et al. \cite{LaxmiLydia2025IDS}           & 2025 & Deep Learning                & CICIDS2017       & 98.77 & 98.77 & 98.77 & 98.76 \\
Ahmed et al. \cite{Ahmed2025MDFuzzy}                 & 2025 & Random Forest                &  UNSW-NB15         & 99.50 & 98.0  & 97.0  & 96.0 \\

Alhayan et al. \cite{Alhayan2025DeepEnsemble}                 & 2025 & Ensemble of Deep Learning               &  UNSW-NB15         & 99.77 & 98.6  & 98.6  & 98.86 \\

\midrule
\textbf{Our Work}                                     & \textbf{2025} & \textbf{CANGuard (Proposed)} & \textbf{CICIoV2024} & \textbf{99.89} & 99.89 & \textbf{99.89} & \textbf{99.89} \\
\bottomrule
\end{tabularx}
\end{table*}

\label{sec:result_discussion}
We perform comprehensive ablation experiments to validate the effectiveness of each element in our CNN-GRU-Attention architecture, followed by comparison with state-of-the-art methods. The findings illustrate the greater efficacy of our proposed method across multiple evaluation scenarios.

\subsection{Comparative Analysis with State-of-the-Art}

Table~\ref{table:comparison-sota} presents a comprehensive comparison between the proposed ConvGRU-based model and several state-of-the-art intrusion detection approaches evaluated on various benchmark datasets. Our model was evaluated on the CICIoV2024 dataset and attained strong results, with an accuracy, precision, recall, and F1-score of 99.89\%. This represents a significant improvement over existing methods. The Deep Neural Network methodology employed by Neto et al.~\cite{Neto2024CICIoV2024:} attained an accuracy of 96\% and an F1-score of 78\%, while traditional methods like Logistic Regression and AdaBoost yielded even lower performance, with F1-scores of 49\% and 51\%, respectively. The consistent and substantial gains across all evaluation criteria highlight the efficacy of the ConvGRU architecture in capturing temporal dependencies and complex intrusion patterns in IoV traffic.

When compared with other models applied to different datasets, such as the 99.77\% accuracy achieved by Alhayan et al.~\cite{Alhayan2025DeepEnsemble} on the UNSW-NB15 \cite{Reaj2025xFEBERT} and the 98.77\% accuracy reported by LaxmiLydia et al.~\cite{LaxmiLydia2025IDS} on CICIDS2017 \cite{haque2021erp}, our proposed model still demonstrates superior performance. It is important to note that direct comparisons across different datasets should be made with caution \cite{taslimul2026role}, as the difficulty levels and data distributions may vary. Nevertheless, the proposed model consistently achieves consistently high scores across all metrics, indicating both high detection accuracy and robustness against false positives and false negatives. Such balanced performance is essential in the field of intrusion detection for Internet of Vehicles (IoV), where both undetected threats and false alarms can have severe implications. Overall, the ConvGRU-based approach provides a strong empirical baseline on the CICIoV2024 dataset and demonstrates significant efficacy for real-time, reliable intrusion detection in intelligent transportation environments.

\subsection{Ablation Study and Discussion}

We conducted Component-wise Performance Analysis in the ablation study. Table~\ref{tab:ablation_results} summarizes the impact of individual components (CNN, GRU, and the attention mechanism) on overall model performance.

The \textit{CNN Only} model achieves solid performance (Accuracy: 0.9933), demonstrating its effectiveness in extracting local spatial patterns. However, its inability to model temporal dependencies limits its generalization for sequential data \cite{taslimul2026erm}, slightly constraining recall and F1-score. The \textit{GRU Only} model achieves slightly lower accuracy (0.9904) but notably higher precision (0.9986). This suggests that GRU effectively captures temporal dependencies and reduces false positives. However, the absence of spatial feature extraction leads to lower recall and an overall drop in balance across metrics. Combining both modules in the \textit{CNN+GRU (No Attn.)} model yields a significant performance improvement (Accuracy: 0.9986). This confirms the complementary strengths of CNN and GRU, where CNN captures spatial hierarchies and GRU models temporal dynamics, resulting in richer and more robust feature representations. The full architecture, \textit{CNN+GRU+Attn.}, delivers superior results across all evaluation measures, reaching an accuracy of 0.9989. The attention mechanism enables the model to dynamically prioritize the most informative features within the sequence, enhancing the quality of learned representations. Notably, the addition of attention yields a 2.45\% improvement in accuracy over the CNN+GRU baseline \cite{jawadul2026health}, while also boosting precision, recall, and F1-score. This demonstrates the critical role of attention in refining feature importance and enhancing the model's focus during prediction.

To mitigate overfitting, the proposed model was trained with dropout, L2 regularization, batch normalization, early stopping, and gradient clipping. BorderlineSMOTE was applied only on the training set to handle class imbalance, while the test set was kept untouched to ensure unbiased evaluation. The close agreement between training and validation performance indicates that the high accuracy is due to effective feature learning rather than memorization.

\begin{figure}
    \centering
    \includegraphics[width=1\linewidth]{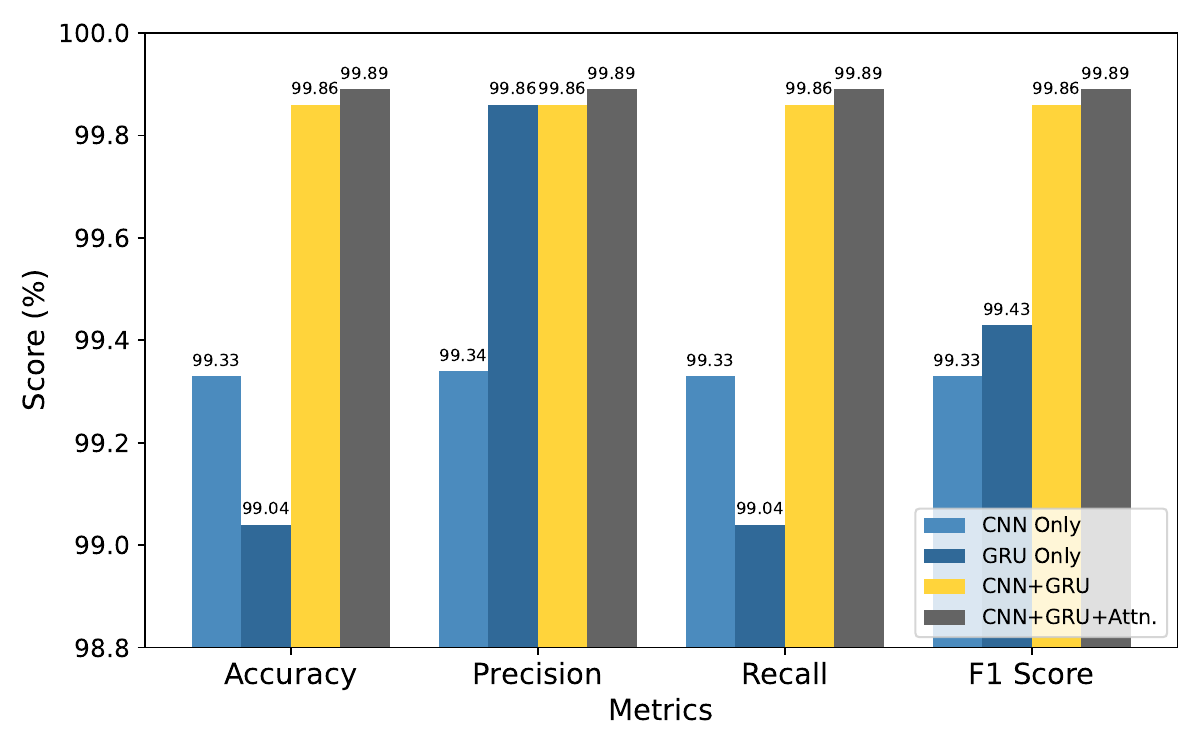}
    \caption{Performance comparison of models in the ablation study, evaluated across different key metrics.}
    \label{fig:placeholder}
\end{figure}
\begin{table}[htbp]
\centering
\caption{Ablation Study Results with Component Indicators}
\label{tab:ablation_results}
\renewcommand{\arraystretch}{1.2}
\begin{tabularx}{\linewidth}{
  >{\centering\arraybackslash}p{0.2cm} 
  >{\centering\arraybackslash}p{0.5cm} 
  >{\centering\arraybackslash}p{0.5cm} 
  >{\centering\arraybackslash}p{0.5cm} 
  >{\centering\arraybackslash}p{1cm}       
  >{\centering\arraybackslash}p{1cm}       
  >{\centering\arraybackslash}p{0.8cm}        
  >{\centering\arraybackslash}p{1.2cm}        
}
\toprule
\textbf{No.} & \textbf{CNN} & \textbf{GRU} & \textbf{Attn.} & \textbf{Accuracy} & \textbf{Precision} & \textbf{Recall} & \textbf{F1-Score} \\
\midrule
1 & \checkmark &             &             & 0.9933 & 0.9934 & 0.9933 & 0.9933 \\
2 &           & \checkmark  &             & 0.9904 & 0.9986 & 0.9904 & 0.9943 \\
3 & \checkmark & \checkmark  &             & 0.9986 & 0.9986 & 0.9986 & 0.9986 \\
\hline
\textbf{4} & \checkmark & \checkmark  & \checkmark  & \textbf{0.9989} & \textbf{0.9989} & \textbf{0.9989} & \textbf{0.9989} \\
\bottomrule
\end{tabularx}
\end{table}

\subsection{Feature Importance Analysis using SHAP Values}
To understand feature contributions to the CNN-GRU-Attention model's predictions, we performed a SHAP (SHapley Additive exPlanations) analysis, which helps interpret the influence of individual features, ensuring model transparency and reliability.

\begin{figure}
    \centering
    \includegraphics[width=1\linewidth]{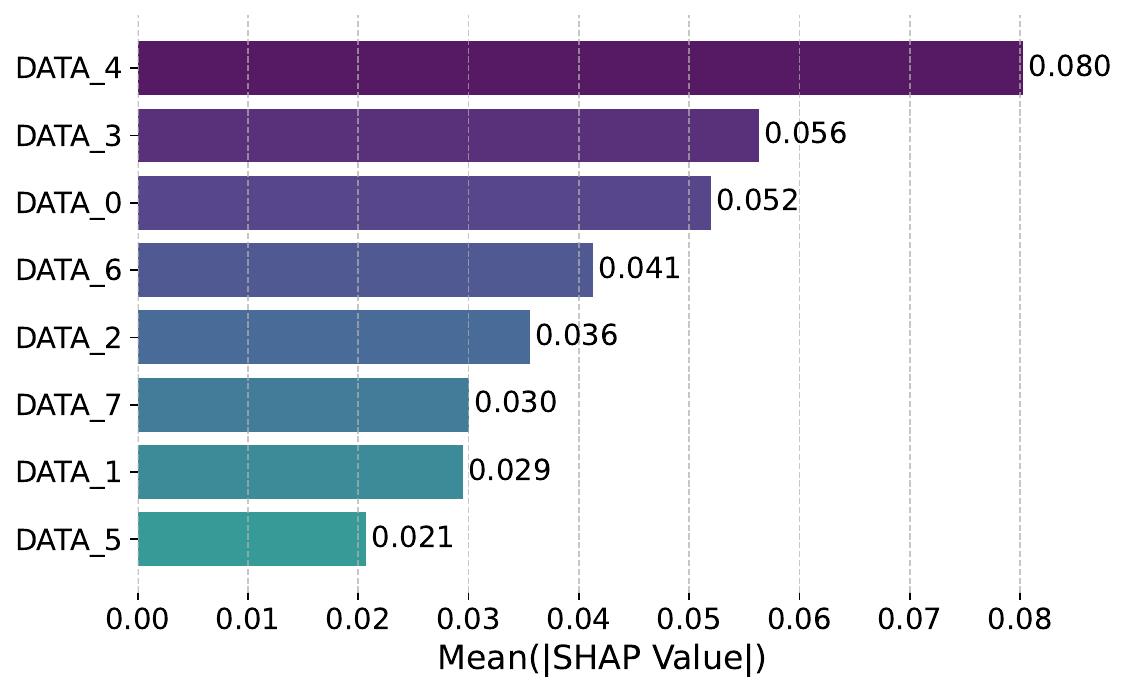}
    \caption{Bar plot of mean absolute SHAP values for DATA\_0 to DATA\_7 features, representing bytes of data transmitted over the CAN bus.}
    \label{fig:feature-importance}
\end{figure}

Figure~\ref{fig:feature-importance} presents the mean absolute SHAP values for the DATA\_0 to DATA\_7 features, which correspond to the bytes of data transmitted over the CAN bus. These features are ranked based on their impact on the capability of the model to distinguish normal traffic from malicious traffic. The analysis reveals that certain data bytes, particularly {DATA\_4} and {DATA\_5}, contribute the most to the model’s predictions, indicating that these payload positions carry significant information for distinguishing benign from malicious traffic. This aligns with the behavior of spoofing attacks in the CICIoV2024 dataset, where specific bytes in the CAN payload are manipulated to alter signals such as speed, RPM, or steering. In contrast, {DATA\_0}, {DATA\_1}, and {DATA\_2} show lower importance, suggesting a lesser impact on the decision boundary for DoS and spoofing detection.

\

\section{Conclusion}
\label{sec:conclusion}
This work introduces CANGuard, a reliable and explainable deep learning framework designed for intrusion detection in Internet of Vehicles (IoV) systems. By integrating CNN for spatial feature learning, GRU for temporal pattern recognition, and an attention mechanism for feature prioritization, the proposed hybrid architecture effectively identifies both Denial-of-Service (DoS) and spoofing attacks on the CAN bus. Experimental results on the CICIoV2024 dataset demonstrate that CANGuard achieves consistently high detection performance across all key metrics, surpassing traditional and deep learning baselines. A detailed ablation study further validates the contributions of each architectural component. To improve the model’s explainability, a SHAP (SHapley Additive exPlanations) analysis was conducted, providing valuable insights into the feature importance in predicting intrusion events. This work contributes a practical, scalable, and explainable solution to securing vehicular networks, and establishes a strong foundation for future advancements in intelligent transportation cybersecurity.

While the results are promising, this study is limited to offline evaluation on a single benchmark dataset and does not include real-time CAN deployment or adversarial robustness analysis. In future work, CANGuard will be extended to cross-dataset evaluations, online CAN bus monitoring, and adversarial settings to further assess its robustness in real-world IoV environments.

\bibliographystyle{IEEEtran}
\bibliography{references}

\vspace{12pt}
\end{document}